# PARALLEL RANDOM SEARCH ALGORITHM OF CONSTRAINED PSEUDO-BOOLEAN OPTIMIZATION FOR SOME DISTINCTIVE LARGE-SCALE PROBLEMS
### L.Kazakovtsev

The the random search methods are implemented to solve the wide variety of the large-scale discrete optimization problems when the implementation of the exact solution approaches is impossible due to large computational demands. Initially designed for unconstrained optimization, the variant probabilities method (MIVER) [1, 3] allows to find the approximate solution of different linear and non-linear pseudo-Boolean optimization problems with constraints [2, 4]. Although, in case of the large-scale problems, the computational demands are also very high and the precision of the result depends on the spent time. In case of the constrained optimization problem, even the search of any permissibly solution may take very large computational resources.

The rapid development of the parallel processor systems which are often implemented even in the computer systems designed for home use allows to reduce significantly the time spent to find the acceptable solution with speed-up close to ideal. The development of the algorithm able to use the capabilities of the cluster systems also allows to use the computational resources of the whole cluster with rather high parallel efficiency and minimum additional load of the communication channels.

In this paper, we consider an approach to the parallelizing of the algorithms realizing the modified variant probability method with adaptation and partial roll-back procedure for constrained pseudo-Boolean optimization problems. Existing optimization algorithms are adapted for the systems with shared memory (OpenMP library used) and cluster systems (PVM library). The parallel efficiency is estimated for the large-scale non-linear pseudo-Boolean optimization problems with linear constraints.

## 1. Constrained pseudo-Boolean optimization problems and random search methods

Often, the requirement of discreteness of the variables takes place (sometimes implicitly) in the technical and economic problems having practical importance, due to standards and production potentialities. We are allowed to formulate all the practical discrete optimization problems as problems of pseudo-Boolean function optimization [1,3,7]. That is why, the methods of the pseudo-Boolean optimization have very wide range of application.

Most practical significant pseudo-Boolean optimization problems are constrained optimization problems.

Most exact solution discrete optimization approaches are based on branch-and-bound method (tree search). Applications of tree search are found in such areas as routing, theorem proving, game theory, combinatorial optimization, artificial intelligence. The classical examples of the problems solved with branch-and-bound method are the integer knapsack problem, the travelling salesman problem etc. [1,2,3,7]. Unfortunately, most of the interesting problems for which the branch-and-bound method provides a viable solution approach are in the complexity class NP-hard and require searching a tree of the exponential size. The process of the further development of the branch-and-bound method creates a wide variety of methods [18] and approaches which reduce the complexity of the solved problems by taking into consideration the distinctive features of the solved problems and searched trees. But, in general, the complexity of the solved problems stays in the NP-hard class. The algorithms based on branch-and-bound methods are easy to be paralellized and their realization on the parallel systems reduces significantly the time spent to the solution. But even the capabilities of the large parallel systems do not allow us to solve the very wide range of the large-scale pseudo-Boolean optimization problems in acceptable time without the significant simplification of the initial problem.

The random search methods are heuristic methods. They do not guarantee the exact solution but random search methods are statistically optimal. I.e. the percent of the problems solved "almost optimal" grows with the increase of the problem dimension [1].

Let's consider the problem:

$$F_M(X) = \sum_{i=1}^{N} \sum_{j=1}^{V} a_{ij} x_{ij} \to max;$$  (1)

$$F_Q(X) \to max;$$  (2)

with constraints:

$$\begin{cases} \sum_{i=1}^{N} \sum_{j=1}^{V} b_{ij1} x_{ij} \leqslant B_1; \\ \sum_{i=1}^{N} \sum_{j=1}^{V} b_{ij2} x_{ij} \leqslant B_2; \\ \quad \dots \\ \sum_{i=1}^{N} \sum_{j=1}^{V} b_{ijN_{Constr}} x_{ij} \leqslant B_{N_{Constr}}; \end{cases}$$  (3)

$$\sum_{j=1}^{V} x_{ij} \leqslant 1 \, \forall \, 1 \leqslant i \leqslant N.$$  (4)

Here, $F_M(X)$ is the first objective function (first criterion), usually linear, $x_{ij}$ are the boolean variables, $F_Q(X)$ is the second objective function (second criterion), often non-linear, $a_{ij}$, $b_{ijk}$ and $B_1$, $B_2, ..., B_{Nconstr}$ are some constants, $N$ is the quantity of chosen elements (goods, units, channels etc.), $V$ is the number of variants for each of the chosen elements.

The problem is to select some set from $N$ possible elements where each of the elements has $V$ variants. The variable $x_{ij}$ is set to 1 if the the resulting set includes $i$-th element in its $j$-th variant.

Our problem has 2 criteria. First one is the typical criterion of the knapsack problem. It is an quantitative estimation of the efficiency of the optimized system. For the economic systems, it is an monetary estimation as usual, the forecast of the profit or losses as an example. We used the net present value as a criterion for the problem of assortment planning[24,10]. For technical systems, it is usually easy to find the similar criterion. For example, in [12] we choose the total cost of the traffic spent during a period of time for the problem of optimization of routing of the telecommunication channels:

$$P(X) = \sum_{i=1}^{N} \sum_{j=1}^{V} P_j(v_{j0} + v_i^* q x_{ij} p_{ij}) \to min.$$  (5)

Here, $N$ is the number of the traffic classes, $V$ is the quantity of interfaces, $P_j$ is the function of the cost of the traffic routed through the $j$-th interface per a month, its argument is the prognosis of the volume of the spent traffic, $v_{j0}$ is the volume of the traffic that has been already spent by the $j$-th interface, $v*_i$ is the average value of the used traffic volume of the $i$-th class calculated per a day (or another time scale), q is the quantity of days (or another time intervals) that remains till the end of this month , $p_{ij}$ =0 if the traffic of the $i$-th class routed through the $j$-th interface is free of charge else $p_{ij}$=1, $x_{ij}$ is a Boolean variable which determines the routing of the $i$-th class of the traffic through the *j*-th interface.

The second criterion indicates the qualitative estimation of the solution. For example, in the problem of optimization of the traffic balance, we can choose the part of the total transfer rate of all channels which stays idle when the total load of the system is high :

$$F_Q(X) = \frac{\sum_{j=1}^{V} min\left\{\sum_{i=1}^{N} a_i x_{ij}, c_j\right\}}{\sum_{j=1}^{n} c_j} \to min. \tag{6}$$

Here, $c_i$ is the capacity of the *i*-th channel, $a_i$ is the average volume of the i-th class traffic per one second calculated in such periods when the load of all channels or even one of them is high (more than some determined percent) [12,14].

In our case, the second criterion is non-linear.

Our optimization problem has 2 criteria, non-linear in general. The prevailing methods of transformation of the multicriterial optimization problem into the 1-criterion one are the convolution method and the Pareto method. In the Pareto method, one of the criteria is transformed into the constraint. The problem with the single criterion is solved many times with different values of the right side of this constraint. The decision maker then selects one of the decisions called Pareto-optimal decisions [8,5,3].

The second way of transformation into the one-criterion optimization problem is the method of convolution. In the simplest form of this method, we replace all the criteria with the sum of the criteria multiplied by some coefficients. In this case, it is very difficult to choose the proper values of the coefficients. We offer the approach of the multiplicative convolution. In this case, we replace all the criteria with their production. It means that the decrease of the qualitative criterion by 1% is equal to the same decrease of the quantitative (monetary) criterion:

$$F(X) = F_M(X) F_Q(X) \quad . \tag{7}$$

This resulting function *F(X)* is always non-linear.

Selection problems such as one above are formulated so that the Boolean variables form a matrix with *N*x*V* dimension. The algorithms [10,12] for the problems of that type are also realized so that the variables form a matrix or even more complex 3-dimensional structure because in this form, it is less difficult to take into consideration the constraints (4) and the results of the algorithm are able to be easy interpreted (the value of 1 in the position $x_{ij}$ means that we select the *i*-th element of the selection in its *j*-th variant). But this form of variable names is not usual for the optimization literature. The matrix of variables can be easy transformed into the vector and our problem can be formulated as:

$$F(X) = \sum_{i=1}^{N \cdot V} a_i x_i F_Q(X) \to max; \tag{8}$$

$$\sum_{i=1}^{N \cdot V} b_{ik} x_i \leqslant B_k \ \forall \ 1 \leqslant k \leqslant N_{constr} \quad ; \tag{9}$$

$$\sum_{i=l \cdot (V-1)+1}^{l \cdot V +1} x_i \leqslant 1 \ \forall \ 1 \leqslant l \leqslant N \ ; \quad . \tag{10}$$

The problem like this is allowed to be solved with the determined algorithm (branch-and-bound method) only if its dimension does not exceed hundreds variables even is the realization of the branch-and-bound algorithm takes into consideration all the peculiarities of the problem and excludes most of the search tree at the first steps. The real large-scale problems have sometimes millions variables. For example, the problem of assortment planning of the wholesale or retail trade company may include the selection of thousands goods names which can be obtained from hundreds suppliers and have 3-10 variants of retail price. In general, problems of such kind are able to be solved only with random search algorithms.

Being initially designed to solve the unconstrained optimization problems, the method of variant probabilities (MIVER) is the random search method organized by the following common scheme [1,3].

1. $k$=0, the starting values of the probabilities $P_k$={$p_{k1}$, $p_{k2}$, ... , $p_{kN}$} are assigned where $p_{kj}$=$P${$x_j$=1}. The starting probabilities is a very significant question for the constrained optimization problems (see section 2).
2. With probabilities defined by the vector $P_k$, there are generated a set of the independent random points $X_{ki}$.
3. The function values in these points are calculated: $F(X_{ki})$.
4. Some function values from the set $F(X_{ki})$ and corresponding points $X_{ki}$ are picked out (for example, point with maximum and minimum values).
5. On the basis of results in item 4, vector $P_k$ is modified.
6. k=k+1, if k<R then go to 2. This is the stop condition which may differ depending on the algorithm realization.
7. Otherwise, stop and

$$F_{min}(X) = \min_{k=1,\dots,R} \left\{ \min_{i=1,\dots,N} f(X_{ki}) \right\} \quad . \tag{11}$$

To be implemented for the problems like (8-10), this method has to be modified. The modified version of the variant probability method, offered in [10,11,12] allows us to solve large-scale problems with dimensions up to millions of the Boolean variables.

In case of the large-scale problems, even the calculation of the linear objective function takes significant computational resources. In the practical problems, the number of constrains grows with the increase of the problem dimension. So, the calculation of the objective function and the constraints is a very large computational problem if it is repeated many times. For the tasks with the large quantity of variables , often we cannot acquire the solution in the acceptable period of time.

That is why, the distribution of the computational tasks between the parallel processors or cluster nodes seem to be very actual.

For the practical technical and economic problems which need the variant probability method for their solution, the realization of the algorithm takes into account the particularities of each concrete problem. Such algorithms are often realized as the computer programs specially designed for the concrete kind of problems.

In this paper, we do not consider greedy search algorithms (which are deterministic or also randomized [19]) though they are often used to improve the results of the random search methods as the final step of them.

Also, we do not consider the genetic algorithms which parallelization methods are offered by many authors [16,17,21] though some approaches offered for genetic algorithms may be implemented for radnom search algorithms parallelization.

Here, we offer an approach of adaptation of the existing programs realizing the random search methods of constrained pseudo-Boolean optimization to be implemented in the parallel systems.

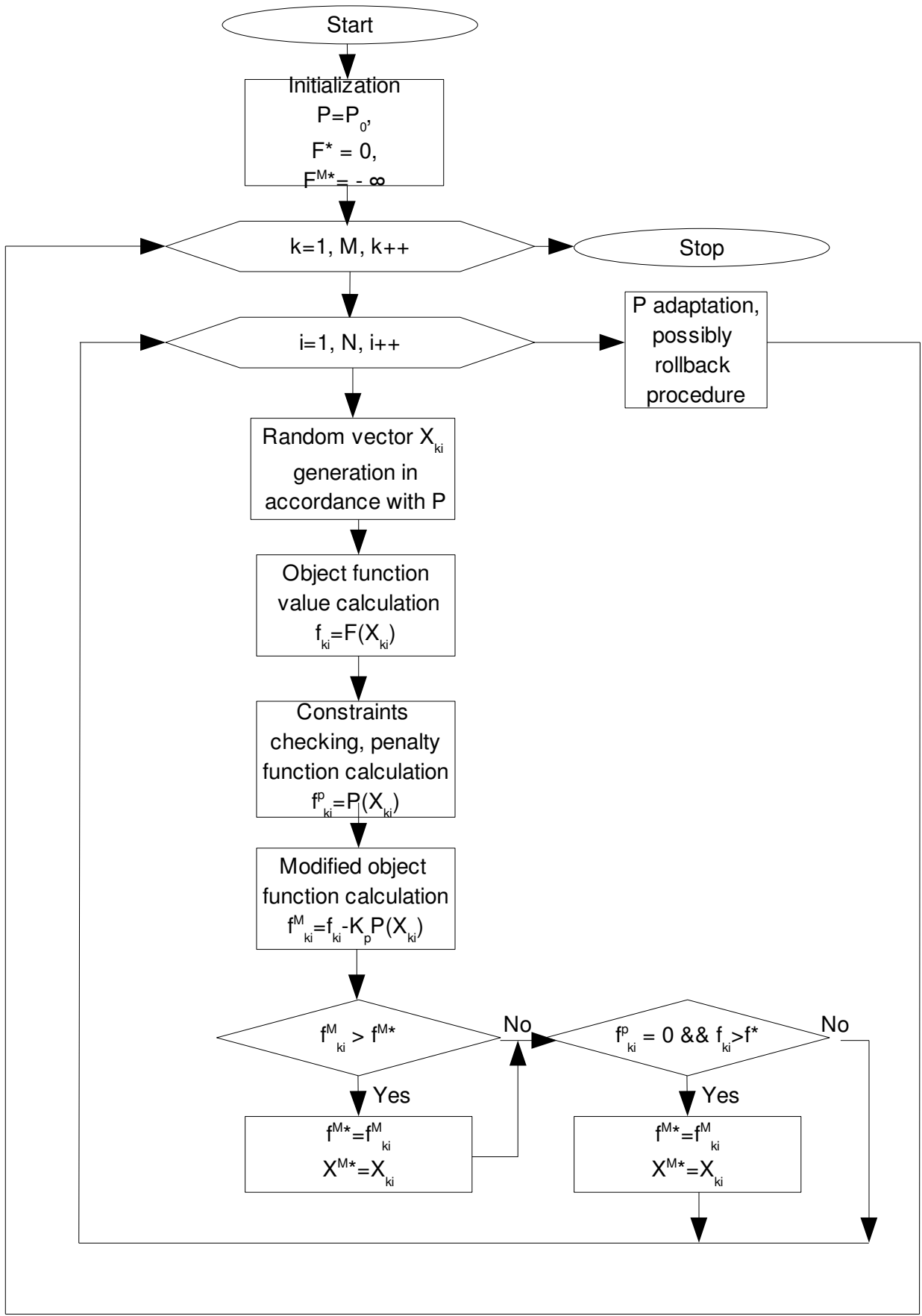

**Figure 1**. Serial algorithm

## 2. Serial realization of the algorithm

The scheme of the algorithm for the serial systems is shown in Figure 1.

At the step of initialization, all the variables p (components of the probability vector P) are set to their initial values ($0 < p < 1$ $\forall$ p $\underline{in}$ P) . Usually, all components are set to 0.5 but, since our algorithm realizes the method of constrained optimization, the initial value of the probability variables is very significant. At each further step of our algorithm, in correspondence with the probability vector, we generate the vectors X of optimized boolean variables. The components of probability vector P determine the probability of generating the value of 1 for the corresponding component of the vector X. In our case of constrained optimization, the large values of vector P components generate the values of X which are out of the area of allowed solutions due to the constraints (9,10). Due to the constraints (10), the optimal initial values of the vector P components do not exceed 1/(L+1) [12].

We set the initial values of the vector P to 1/(L+1) but we have to reduce this value if several starts of our algorithm give us no results in the allowed solutions area. This peculiarity is not illustrated in the picture 1 for the simplicity.

In first cycle (k=1 to M), M determines the maximum number of steps of the vector P adaptation. Also, we can use the maximum run time as the stop condition. In some cases, it is reasonable to use the maximum number of steps which do not give us the result better than previous one as the stop condition. In figure 1, we illustrate the simplest case when we use the maximum number of adaptations as the stop condition.

In the second cycle (i=1,N), we generate the set of N vecors $X_{ki}$ in acordance with the probability vector P. Then, the objective function is calculated for each $X_{ki}$. Also, we calculate the values of the left parts of our constraints, and introduce the second objective function $f^p$ (penalty function):

$$f_{ki} = F_P(X) = C_{PENALTY} \sum_{k=1}^{N_{CONSTR}} F_{Pk}(X); \tag{12}$$

$$f_{ki}^P = F_{Pk}(X) = \left\{ \begin{array}{l} 0, \sum_{i=1}^{N \cdot V} b_{ik} \leqslant B_k \\ \dfrac{\sum_{i=1}^{N \cdot V} b_{ik}}{B_k}, \sum_{i=1}^{N \cdot V} b_{ik} > \end{array} \right\} \tag{13}$$

Here, $C_{PENALTY}$ is some coefficient. If some estimation of the maximum value of the objective function is available, we can use it as the value of the coefficient $C_{PENALTY}$. For linear objective functions:

$$C_{PENALTY} = \sum_{i=1}^{N \cdot V} |a_i| \quad . \tag{14}$$

The modified objective function $f^M$ for our optimization task is the sum of the initial objective function and penalty function.

When all the values of $f_{ki}$ and $f^M_{ki}$ are calculated, we choose the best (maximum) value for $f^M_{ki}$ and, if there were the variants of the vector $X_{ki}$ which satisfy all the constraints ($f^p_{ki}$=0) then we choose the maximum value of the objective function $f_{ki}$ for that variants of the vector $X_{ki}$.

The essential for our algorithm is the step of adaptation of the vector $P$. In the original MIVER method, the additive adaptation is used:

$$p_{kj} = min(1, p_{k-1\,j} + d_k), \text{ if } x^{max}_{kj} = 1 \text{ and } X^{min}_{kj} = 0,$$
$$p_{kj} = max(0, p_{k-1\,j} - d_k), \text{ if } x^{max}_{kj} = 0 \text{ and } X^{min}_{kj} = 1. \tag{15}$$

Here, $p_{kj}$ is the $j$-th component of the vector $P$ at the $k$-th random vector set generation, $d_k$ is the addition value at the $k$-th step. It may be constant or decrease. For example,

$$d_k = 0.1/k.$$

$x^{max}_{kj}$ is the j-th component of the vector $X^{max}_k$ which gives the maximum value of the modified function $f^M(X)$ at the $k$-th generaton, $x^{min}_{kj}$ is the component of the vector giving the minimum value of the modified function.

In our case of the constrained optimization, this method of adaptation needs to be corrected. The solution of different practical tasks shows the best result if we use the multiplicative adaptation with rollback [3,10]. In that case, the components of the vector $P$ are never set to the value of 0 or 1 which may cause that all the further generations of the $X$ vector have the same value of the corresponding component which does not give the allowed solution due to constrains (10).

$$p_{k,j} = \begin{cases} p_{(k-1),j} \cdot d \,, & x^{max}_{kj} = 1 \wedge x^{min}_{kj} = 0 \wedge p_{(k-1),j} < 0.5 \,, \\ 1 - \dfrac{(1 - p_{(k-1),j})}{d} \,, & x^{max}_{kj} = 1 \wedge x^{min}_{kj} = 0 \wedge p_{(k-1),j} \geqslant 0.5 \,, \\ \dfrac{p_{(k-1),j}}{d} \,, & x^{max}_{kj} = 0 \wedge x^{min}_{kj} = 1 \wedge p_{(k-1),j} < 0.5 \,, \\ 1 - (1 - p_{(k-1),j}) \cdot d \,, & x^{max}_{kj} = 0 \wedge x^{min}_{kj} = 1 \wedge p_{(k-1),j} \geqslant 0.5 \,, \\ p_{(k-1),j} \,, & x^{max}_{kj} = x^{min}_{kj} \,. \end{cases} \tag{16}$$

Here, $d$ is the adaptation coefficient. In case of multiplicative adaptation, it does not depend on the step number $k$. In this case, the absolute value of adaptation step depends on the corresponding value of $p_{kj}$.

In some modifications of MIVER method, the rollback procedure is performed. After several steps, the values of P vector elements are close to 0 or 1 and the decrease of the adaptation step ($d$) results in generation of the similar vector $X$ exemplars which correspond to some local maximum. The rollback procedure is helpful to avoid that situation. It sets the values of $P$ vector and its adaptation step $d_k$ to initial (or other) values. The conditions of the rollback procedure start may differ. In simplest case, it starts after several steps. Also, it may be performed after several steps which do not give us the objective function value which is significantly better than previous ones (the increase of the objective function does not exceed $\Delta_f$):

$$f^M_k - f^M_{k-m} < \Delta_f. \tag{17}$$

Here, $f^M_k$ is the maximum value of the modified objective function after $k$ steps, $f^M_{k-m}$ is its maximum value after $k$-$m$ steps.

In case of constrained optimization, the presence of the rollback procedure is more important than in case of the unconstrained optimization due to the complex shape of the admissible solutions area [10]. For our type of the problems, we use special kind of rollback methods which have shown the good results for different problem types [1]. In simplest case, the probability vector $P$ is set to ist initial value and its information of the objective function behaviour is cleared. The better results are demonstrated with methods of partial rollback which change some part of $P$ vector components or change all the components so that their new values depend on previous results. We can use the following rollback formula:

$$p_{kj} = (p_{k-1\,j} + q_k p_0)/(1+q_k), \text{ if } p_{k-1\,j} < p_0. \tag{18}$$

Here, $p_0$ is the initial value of the probability. The coefficient $q_k$ may be constant or vary depending on the results of previous steps. For example, it may depend on the quantity of the steps which do not improve the maximum result ($s_m$).

$$q_k = w\,/\,s_m.$$

The weight coefficient $w$ has to be chosen experimentally. It depends on the frequency of the implementation of the rollback procedure. If this coefficient is small enough, the rollback procedure can be implemented at each step. For the optimization problems with thousands variables, we use the values 0.01 ... 0.05 with partial rollback procedure at each step.

In case of constrained optimization, the choice of the initial value $p_0$ can be very important. In some cases, the incorrect initial value causes the generation of $X$ vector sets that lay out of the allowed solutions area. After several steps, the penalty function minimization process results in adaptation of $P$ vector which allows to generate $X$ vector exemplars that satisfy our constraint conditions. But the rollback procedure returns the probability vector to its initial (usually incorrect) value. Let's consider a simplest example. Let our problem have only one constraint like that:

$$b_1 x_1 + b_2 x_2 + b_3 x_3 + \ldots + b_D x_D < B. \tag{19}$$

Here, $D$ is the dimension of our problem ($D = N\,V$). The elements of the vector $X$ at the first steps are generated so that it is set to the value of 1 with probability $p_0$. The expectation of the left part of (19) is

$$M = p_0 b_1 + p_0 b_2 + p_0 b_3 + \ldots + p_0 b_D = (b_1 + b_2 + b_3 + \ldots + b_D)\,p_0. \tag{20}$$

The maximum values of the objective function are usually achieved at the points which lay near the boundary of the constrained area ($a_1 x_1 + a_2 x_2 + a_3 x_3 + \ldots + a_D x_D \sim = A$ in our case). Therefore, the optimal value of the initial probability is

$$p_0 = B/(b_1 + b_2 + b_3 + \ldots + b_D). \tag{21}$$

In more complex cases, especially if the constraints are non-linear, it is difficult to determine the optimal initial probability value. But the negative effect of the rollback procedure can be reduced by adaptation of the initial value of $p_0$:

$$p_{0\,k} = (p_{1\,k-1} + p_{2\,k-1} + \ldots + p_{D\,k-1})/D. \tag{23}$$

This version of rollback procedure is most actual in case of partial rollback which after performed at each step of our algorithm. The implementation of this kind of adaptation&rollback procedure is described below for the parallel version of the optimization algorithm for systems with no-remote memory access (clusters).

### 3. Parallel version for multiprocessor systems

The adaptation of our algorithm for multiprocessor systems with shared memory can be performed by the parallel generation of the exemplars of the X vector and their estimation. The scheme of that version of our algorithm is shown in figure 2. If our system has $N_P$ processors, the cycle of generation of $N$ exemplars of the vector $X$ can be divided between the processors. Each processor has to generate $N/Np$ exemplars of the vector $X$ and calculate the value of the objective function, left parts of the constraint conditions and calculate the modified objective function values.

The organization of the parallel thread takes significant computational expenses. In [6], they estimate that expenses as 1000 operations of real number division. So, this kind of parallel execution has no sense if the dimension of our optimization problem is not large. But we consider the task with thousands variables. The experiments at 4-processor system with linear 100-dimension problem (105 constraints) show that the parallel version runs 2.8 times faster than the serial one. Often, the dimensions of the problem exceed 1000 and the use of parallel processors gives much better acceleration.

In our case, the serial part of the algorithm includes the comparison of the generated results and the probability vector adaptation with possible rollback procedure. The computational complexity of all that steps depends on the problem dimension linearly and can be performed during a single pass each. Moreover, the choice of best and worst exemplars of generated vectors can be performed by each processor during each loop of generation and estimation. In the serial part of the algorithm, only the maximum and minimum results calculated by different processors are compared. We do not show these details in figure 2 for the simplicity.

To realize the parallel version of our algorithm, the minimum changes in the source code are needed. If the source code of the program realizing the serial version is written in C/C++ or Fortran and it is possible to use OpenMP library with suitable compilers, all we need is to insert one additional line before the 2nd level cycle. An example for Fortran:

```
        do k=1, M
c$omp parallel do
            do i = 1, N
                <Random vector generation>
                <Generated exemplars estimation etc>

            enddo
            <Adaptation, possible rollback>
        enddo
c$omp end parallel do
```

Also, our program must contain special lines in its initial part ($omp for Fortran, #pragma omp for C++) making the compiler use the OpenMP library with needed options.

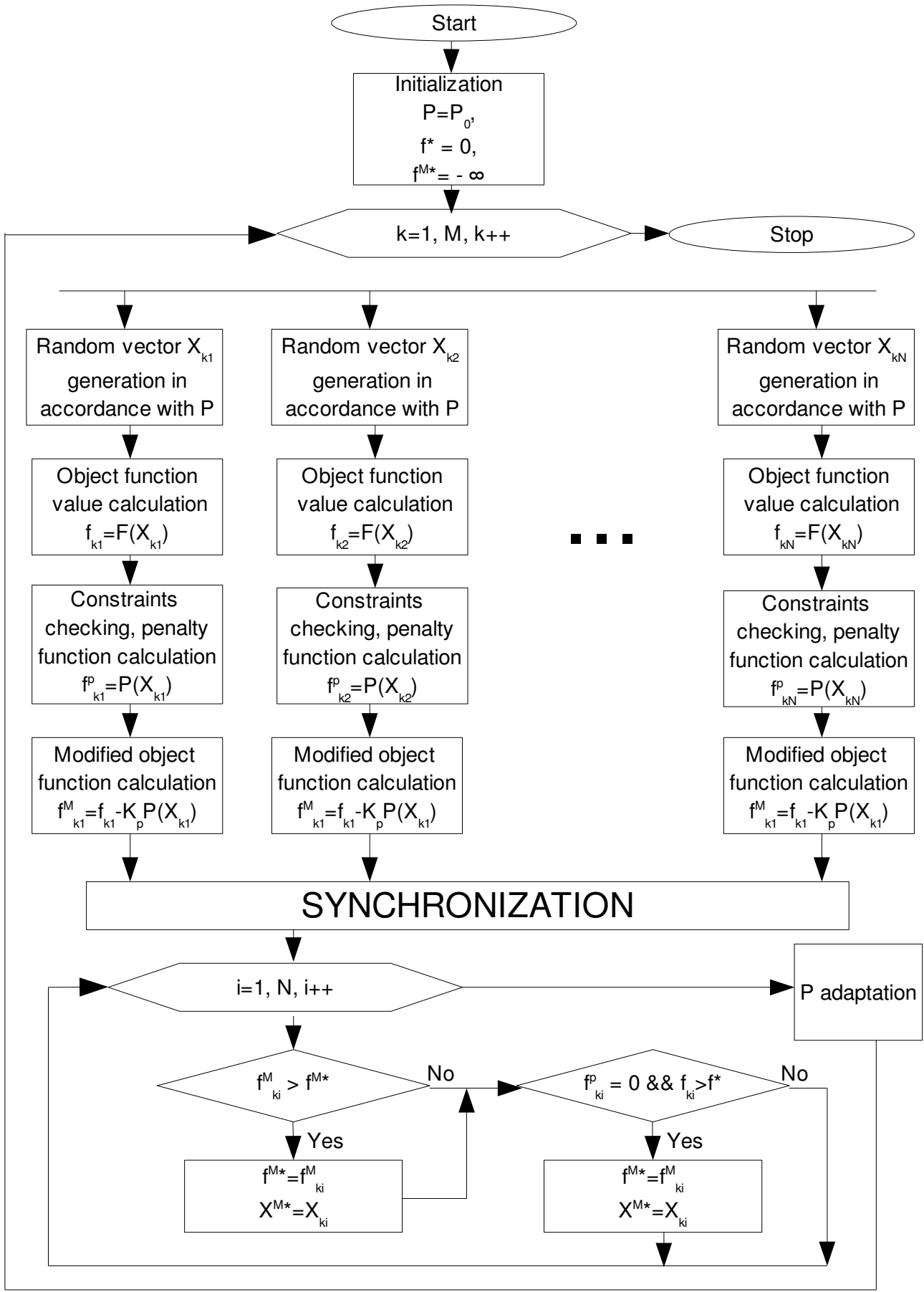

**Figure 2.** Parallel algorithm for systems with shared memory

The loops of the cycle must have no data dependencies. The result of the previous loop of the cycle must not be processed by the following loops.

The loops of the cycle need unequal computational expenses due to the different complexity of the calculation of the modified objective function for exemplars which lay inside and outside the allowed solution area. Also, in case of non-linear objective function and/or constraints, their calculation expenses may differ depending on the arguments contained in X vector. That is why, the dynamic execution mode is needed. In this mode, the loops of the cycle are distributed between processors dynamically. This takes more processor resources to perform the indirect labour of parallel processors load distribution in comparison with the static execution mode which divides the executed cycle into $N/N_p$ cycles and starts their execution simultaneously. In the other hand, due to the random character of X vector exemplars generation, the expectation of the runtime spent to perform each loop is equal. So, in case when each processor performs large quantity of the loops, the processors finish their work almost simultaneously. The practical work takes equal time (1-4% difference) with both types of execution mode if we run the algorithm to solve the linear optimization problems with 100-10000 variables at the 4-processor system.

So, the serial version of the optimization algorithm can be simply transformed into the parallel version for multiprocessor systems which demonstrates the significant reduction of the time spent to solve the problem in comparison with the serial version.

The following listing of a program (fragment) illustrates the modification of the existing fortran program. To paralellize it, we add only 2 lines.

```
ccccccccccccccccccccccccccccccccccccccccccccccc
c   This procedure performs 1 step. It generates the X exemplars, sorts,
c   compares and adapts the probabilities
      subroutine step1(x,p,population,xbest)
c   Number of generated X vectors
      integer population
c   probability vector P and an 2D aray for X vectors
      real p(*)
      integer x(population,10000)
c array to store the X vector which gives the best result
      integer xbest(10000)
      integer nbest,nworst,best0
c Public variables:
c The following public variables contain the max.obj.function value,
c modified object. F-n value, total number of steps and number of steps
c which do not improve the maximum values since last restart
c and variables quantity (dims)
      common /publ/ best_r,best_m_r,nsteps_made,nsteps_noresult,dims
      double precision best_r,best_m_r
      integer nsteps_made,nsteps_noresult,dims
      double precision zn,zp,results(100),zm
      integer k,i,penalties(100)
c Following procedure performs full or partial reset if needed
      call possible_reset(p)
      nsteps_made=nsteps_made+1
      nsteps_noresult=nsteps_noresult+1
c In the parallel cycle we just generate X exemplars and evaluate them
c$OMP PARALLEL DO  PRIVATE(zp,zn,zm)
      do k=1,population-1
c First of all, we generate X vector in accordance with P vector
c and store it to the k-th line of the array X
      call x_generation(x,p,k)
```

```
c We calculate the objective function and the penalty function
        zp=penalty(x,k)
        zn=ff(x,k)
c  modified objective function calculation
        zm=0d0-zp*sum_coef+zn
        results(k)=zm
        if (zp .ne. 0d0) penalties(k)=1
     end do
c$END PARALLEL DO
c   Now, all the Xs are generated. We sort the array of them
      call sorting(x,results,penalties,population,nbest,nworst,best0)
c   NBEST and NWORST contain numbers of best and wotst generated
c   strings of the X array, results(nbest) contains the best
c   generated value, best0 contains the number of string which
c   gives the best admissible solution.
c   Here, we check if the best generated result improves the maximums
      if (results(nbest) .gt. best_r_m) then
           nsteps_noresult=0
           best_r_m=results(nbest)
      end if
      if ((nbest0.gt.0).and.(results(nbest0).gt.best_r)) then
           best_r=results(nbest0)
           do 1113 i=1,dims
1113            xbest(i)=x(nbest0,i)
      end if
c  Here, we call the procedure of adaptation of the vector P
      call adaptation(x,p,nbest,nworst)
      end
```

## 4. Version for multicomputers

The above approach of transformation of the serial optimization algorithm into the parallel one can be realized on the systems with shared memory. The parallel processes generate large information volume (large X vectors exemplars) which is compared and processed in the serial part of the algorithm. In case of systems with distributed memory, this information must be collected at the node which performs the serial part of the algorithm with means of the network interfaces.

In [9,20] authors offer an approach of parallelization of the random search algorithm to be implemented in multicimputer systems. They parallelize the step of generation of the vectors X and their evaluation. This approach is a modification of the method considered in the previous chapter of this paper. But it needs intensive data interchange between the nodes and often synchronization.

The information volume generated by each of the parallel processes is significantly reduced if each of the parallel processes generates series of the X vector exemplars, estimates each of them, and sends as the result only the best and worst exemplars. But the problem of the synchronization for such systems takes so much time and resources that the effect of the parallel execution is not evident. The experiments on the system of 2 computers with 1Gbps network interface (Linux, PVM) show that the effect of the parallel running appears when the dimension of the linear optimization problem of our type exceeds 1000-3000. With less dimension, the effect is negative.

To reduce the transmitted data volume and exclude the synchronization step, we offer another approach.

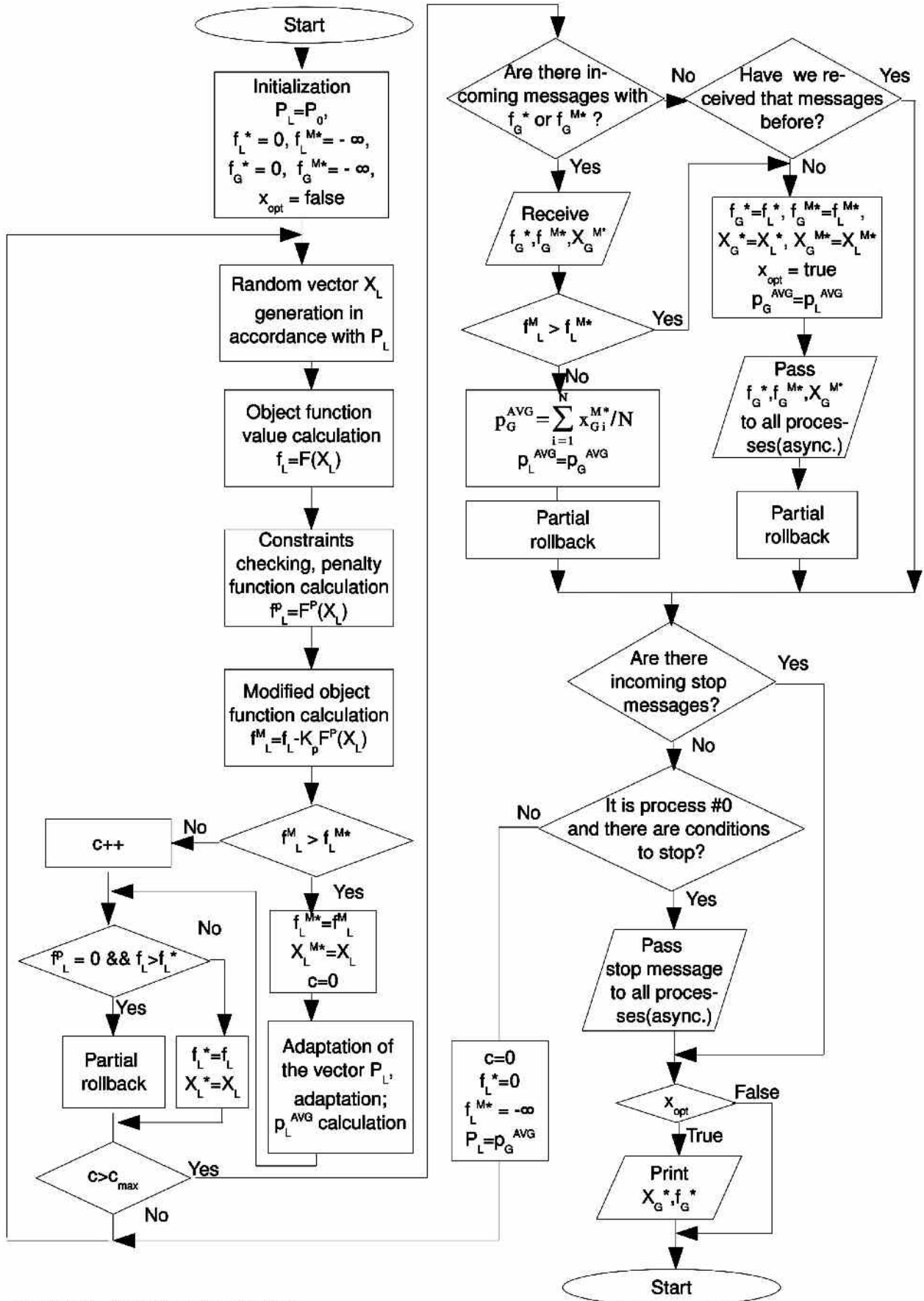

**Figure 3.** Algorithm for clusters

The processes executed at the nodes of the system must be maximum independent. First of all, in case of 2-criterion optimization, we transform our 2-criterion problem into 1-criterion one transforming the $2^{nd}$ criterion into an additional constraint and solve the series of the 1-criterion problems with different values of the right part of that new constraint [13]. These processes are absolutely independent. But then the elements of the Pareto set composed from that solutions are analysed by the decision maker who compares 2-4 different results at one step and determines the possible area for the further search. If we start more parallel processes simultaneously to obtain more results in the Pareto set, most of them are not even requested by the decision maker. In the other hand, the modified optimization problems generated by transformation of the $2^{nd}$ criterion into the constraint, take significantly different computational resources even in the linear case. Some of the generated modified optimization problems may have no decision, others may have trivial decisions. This fact causes large idle time for the nodes solving the modified optimization problems with different parameters.

Another way to organize the parallel execution of the search processes with the maximum independence from each other is to organize multiple starts of the algorithm at all the nodes. The algorithm starts at the nodes with the same or different initial parameters (for example, different initial values of the probability vector elements).These multiple simultaneous starts are performed instead of the rollback procedure in the serial version of the optimization algorithm. Each node starts the cycle of the random $X$ vector generation with probability vector adaptation. If the algorithm does not improve the best objective function value during several steps, the rollback procedure is performed. When all the nodes reach the stop condition, the results of all the nodes are compared to figure out the best one as the final result of the optimization problem.

In its simplest form, the above approach does not need any modification of the implemented software. The serial version of the algorithm runs at all the nodes simultaneously and the results of the nodes are then compared by the operator or by the special node.

But we mentioned above that the full rollback procedure which saves no information collected in the probability vector after its execution is much less effective in case of constrained optimization than the procedure of the partial rollback which corrects the probability vector so that its elements (or only some part of them) are set to the values which are closer to the initial values but the values of the probability vector elements also depend on the values contained in the probability vector before the rollback procedure (18).

The simultaneous execution of the serial version of the algorithm improves the results of the calculations insignificantly due to the similar behaviour of the algorithm at the different nodes. The comparison of the $P$ vector values from different nodes after several steps shows that the difference between them is reduced with each following step and the nodes generate the new exemplars of the $X$ vectors in the very close area. In case of the algorithm with no average probability adaptation, the simultaneous execution of the serial version of the algorithm with different initial probability value gives the results much better than the serial algorithm executed at the single node. But in most cases, the version of the optimization algorithm with the average probability adaptation executed at the single node shows even better results. So, our approach must support the average probability adaptation for the optimization algorithms executed at different nodes (23).

In figure 3, we offer the optimization algorithm for the multicomputer systems with no shared memory performing the information exchange with message passing interface. The most popular examples of that kind of systems are the systems using MPI library (Message Passing Interface) [26] or PVM library (Parallel Virtual Machine). Both libraries are distributed under the GNU/GPL license and included in different versions of the GNU/Linux operating system. We have used the PVM library to build the cluster of nodes with installed Ubuntu Linux v.8.04 but the scheme shown in the figure 1 illustrates the approach to the algorithm development which allows the realization with use of PVM or MPI library or even no special library (in this case, we realize the message passing using the system message passing means). The use of the special library simplifies the process of the parallel execution and organizes the results output but it does not give us any significant speed-up in comparison with the systems which use another means of message passing such as file transfer or common database due to low intensiveness of the message passing of this algorithm. The node which has improved the best result (i.e. maximum value of the objective function) passes the message containing the maximum value and the corresponding $X$ vector value. Other nodes keep silence until one of them improve that result.

In figure 3, the initialization steps of each of the parallel processes include the initialization of the local variables $f_L^*$, $f_L^{M*}$ which contain the best (maximum) values of the objective function and modified objective function achieved by the process and also the pair of variables $f_G^*$, $f_G^{M*}$ which contain the maximum values reached by all the processes. Also, other variables in the figure 3 have corresponding indexes L or G which indicate the character of their values; the variables (scalar or vector ones) containing the values relating to the maximum values achieved by all the processes have the index G, other variables containing the values of the objective functions and generated vectors relating to the local process, have the index L (local). So, the probability vectors P are local, maximum objective function values are both global and local, $X$ vectors generated at each step are local but the vectors which give the maximum values of the objective functions ($X_L^*$, $X_G^*$, $X_L^{M*}$, $X_G^{M*}$) are both local and global.

The steps of X vector generation and objective function and modified objective function calculation coincide with ones of the serial algorithm. The steps of comparison with maximum values and adaptation differ from the serial algorithm. Instead of the cycle rating $N$ exemplars of the vector $X$, we include in our algorithm the counter $c$ which indicates the quantity of steps which do not improve the local maximum value. When this counter exceeds its maximum value $c_{max}$, we check the incoming messages from the other processes. The adaptation step is performed after each step of the X generation. Also, we perform the partial rollback step after each step of X generation. In the partial rollback procedure, the coefficient $q_k$ is set to a rather small value (18).

If the algorithm does not generate the vectors $X$ improving the local maximum values of the objective functions after $c_{max}$ generations, the process tries to communicate to the other processes to check if any other node has generated the vector which has improved the previous global maximum of the objective function or the modified objective function. If so, the process receives that values and the value of the new initial probability value which is calculated by the process having reported the global maximum objective function value as the average value of its $P$ vector value. If the local maximum values $f_G^*$ or $f_G^{M*}$ are better than the received ones then we report our values as the new global maximums. Then, we continue our calculations without the full rollback. Otherwise, we perform the full rollback procedure with initialization of the probability vector $P_L$ with the values of $p_G^{AVG}$.

If the messages sent by the node which has improved the previous maximum value ($f_G^*$ or $f_G^{M*}$) includes also the values of the probability vector $P$, the efficiency of the cluster increases. In that case, the other nodes are allowed to implement the partial rollback procedure after receiving the message and continue their search with the probability vector values close to that of the vector which has generated the best solution.

But the messages containing all the values of $P$ vector are very long. The vector $P$ for the problem with 10000 variables needs 40000 bytes to be sent. To reduce the length of the messages, we implemented another strategy. The node which has achieved the best value, sends this value and the values of the vector $X_G^*$ ($X_G^{M*}$ if the node has improved the maximum value of the modified objective function). For the problem with 40000 variables, the length of the message including $X_G^*$ is about 5000 bytes. Also, the length is decreased if we implement any compression method.

If the node passes the message containing $X_G^*$ then other nodes are allowed to continue their search with new probability vectors which differ from that of the node achieved the best result but our method guarantees that the exemplars of the vector X generated with the new probability vector are close to $X_G^*$. The values of the new probability vector are calculated as

$$p_i = \frac{(C_{corr} + (1 - 2C_{corr}) x_{Gi}^*) p_G^{AVG}}{C_{corr} + (1 - 2C_{corr}) p_G^{AVG}} \quad . \tag{24}$$

Here, $x_{Gi}^*$ is the $i$-th element of the received vector $X_G^*$, $D$ is the dimension of the problem (number of the elements of $X_G^*$), $C_{corr}$ is some small real value, $p_G^{AVG}$ is the average value of the probability vector generated the vector $X_G^*$. If the received message does not contain it then the node which has received the message evaluates it:

$$p_G^{AVG} \approx \frac{\sum_{i=1}^{D} x_{Gi}^*}{D} \quad . \tag{25}$$

Taking into consideration the constraints (10), the value of the constant $C_{corr}$ can be calculated as 0.5/$V$. Also, we are allowed to use $p_G^{AVG}$ as the value of $C_{corr}$.

This approach guarantees that if some element of the vector $X_G^*$ is equal to 1 then the value of the corresponding element of the vector $P$ is close to 1 and the following exemplars of the vector $X$ are generated so that most of them contain 1 in this position. Otherwise, if the element of the vector $X_G^*$ is equal to 0 then the value of the corresponding element of the vector $P$ is close to 0 and the following exemplars of the vector $X$ are generated so that most of them contain 0 in this position. So, the following exemplars of the vector $X$ are close to $X_G^*$ and the nodes continue their search in the area around $X_G^*$. If some of them find a better result, it passes the message. If no, it performs the usual rollback procedure.

So, we realize the parallel multistart which uses the information of the node which has found the best result at previous steps.

The process #0 checks the stop conditions. We use the time elapsed after last report of the global maximum values as the stop condition. The process #0 sends the stop message to the other processes if no processes send reports of improving the global maximum values during the specified period of time. After receiving the stop signal (this step is not shown in figure 3 for the simplicity) the process checks if this process is the one last reported of the global maximum (this fact is indicated in the $x_{opt}$ Boolean variable). If so, the global maximum value and corresponding $X$ vector elements are sent to the output.

The information contained in the messages which are sent between the nodes has very small volume. In fact, we transmit only three real values ($f_G^*$, $f_G^{M*}$ and $p_G^{AVG}$) and a vector of Boolean variables ($X_G^*$). If the bandwidth of the network is very critical, we do not even need to send the value of the $X_G^*$ vector which corresponds to the found maximum objective function value. Its value is kept locally until the stop message is received.

The efficiency of the algorithm is reached due to the simultaneous calculations which are performed in the serial version of the algorithm after each rollbalck. The calculations are interrupted after several steps which do not improve the maximum value if any other node has found the $X$ vector improving the maximum value. Also, the rollback procedure is performed with the new initial values of the probability vector elements which correspond to the average value of elements of the probability vector which has allowed to generate X vector at which the global maximum objective function value is reached. So, the message interface allows the nodes to interrupt the calculations which have no perspective to reach the global maximum and coordinate the process of probability vector adaptation.

## 5. Results

For the parallel random search algorithms, the analytical estimation of the comparative efficiency of the parallel versions in comparison with the corresponding serial algorithms is difficult. The analytical estimations of the efficiency of the random search algorithms are described with the recurrent formulas [28] which need the very concrete information of the objective function (linearity, unimodality etc.) In case of constrained optimization, the analytical estimation of the algorithm efficiency is even more complex problem. The analytical estimation of the parallel version of the algorithm is based on the comparison of the time spent to solve the parallel part of the algorithm (Figure 2) and serial part of the algorithm which includes probability vector adaptation and possible roll-back procedure. The probability vector adaptation complexity depends on the problem dimension linearly. The complexity of the parallel part depends on the dimension of the problem and quantity of the constraints which also depends on the problem dimension linearly for most practical problem types. So, the complexity of the parallel part is proportional to the square of the problem dimension in case of linear objective function. That is why, the comparative efficiency of the parallel algorithm increases with the problem dimension growth. The exact analytical estimation of the comparative efficiency is difficult due to the reasons listed above. In case of the cluster systems, the analytical estimation is even more complex problem.

The experimental estimation of the comparative efficiency of the random search algorithms offered in [15] can be implemented for the constrained optimization case. In our case, we create a set of the test problems. The coefficients of the objective functions and constrains are set by a random generator [7,8,13,25]. The coefficients in the constraints are also generated. The right parts of the constraints are selected so that the problem has admissible solutions. The problems are solved with serial version of the algorithm. In this case, the stop condition is the time spent by the algorithm. For example, algorithm runs 10 minutes. Then, the serial algorithm is implemented K times to solve the same problem. The stop condition in this case is the achievement of the result which is equal or more than the result of the first run. Then, the parallel version of the algorithm runs at the same kind of hardware (in this case, more than one processor or cluster node is used) K times. The stop condition in parallel case is also the achievement of the result exceeding the first achieved objective function value. The total amount of time spent to make K starts of the serial algorithm and K starts of the parallel version are then compared.

In some cases, the decision maker sets the very hard constraints for the problem which has no admissible solution in this case. The result of the algorithm in this case is the establishment of the fact that the admissible solution is not found. We cannot be sure that the problem has no admissible solution because the random search methods do not guarantee the exact solution. In this case, the implementation of the method above is not possible. To estimate the comparable efficiency of the algorithm, we use the condition of the achievement of the penalty function values less than its value achieved with the first start of the algorithm for the problem which has no admissible solution. For that purpose, we generate the problems with no admissible solutions (to generate such a problem, we just set the right part of one constraint condition to 0 when all the coefficients in its left side are positive).

We used one 4-processor computer (4 Xeon processors, 2.6 Ghz) to solve a set of random generated problems with 10000 Boolean variables. To estimate the cluster version of the algorithm, we used a system of 6 nodes with the equivalent hardware (1 Pentium IV, 2.0 GHz processor) connected by a 100 Mbps Ethernet network with configured PVM cluster.

The results for the parallel processor system with shared memory shows the parallel efficiency value 0.97. It means that the parallel system with 4 processors achieves the same results 3,86 times faster than the serial one.

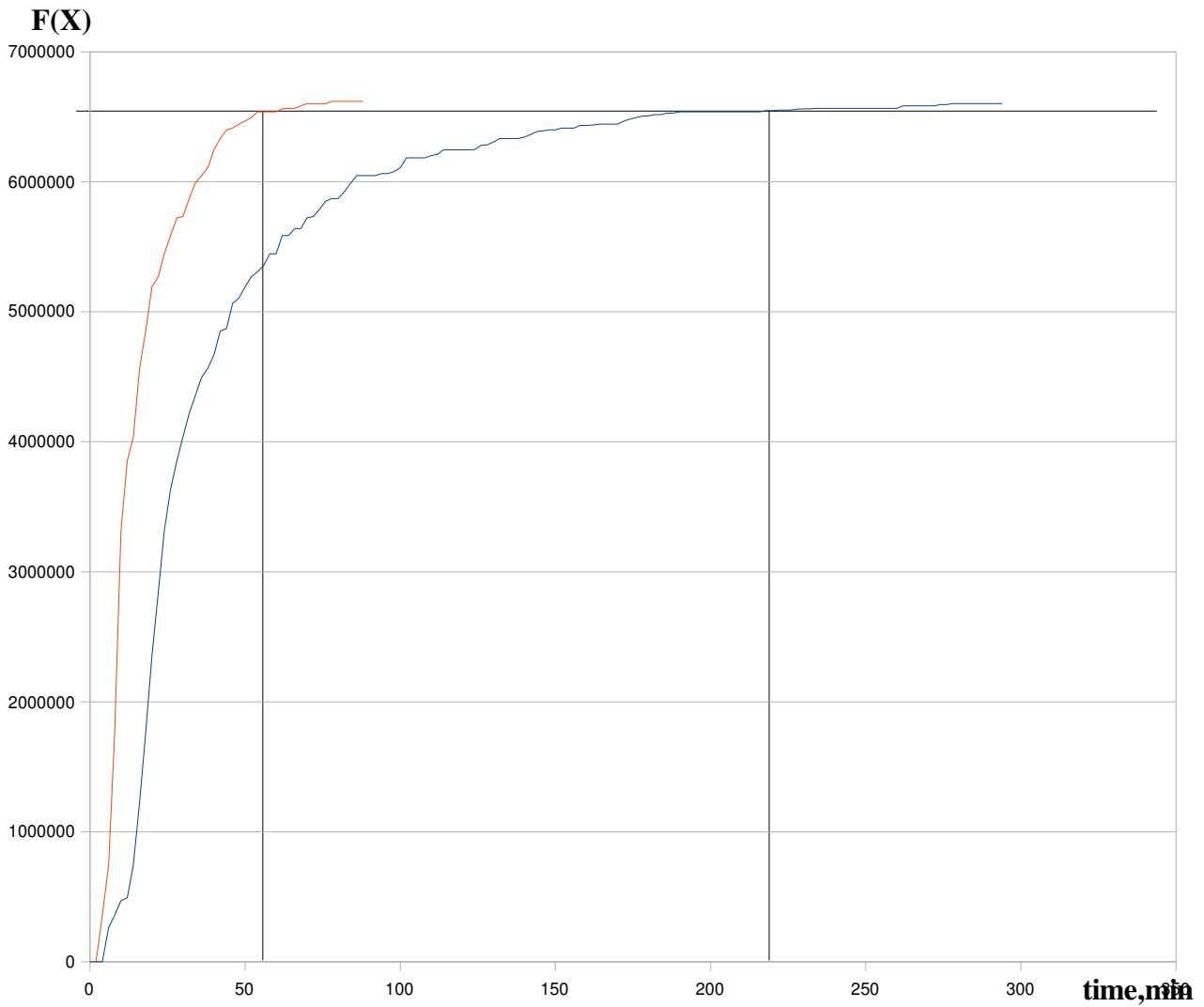

**Figure 4. Comparison of the results of serial and parallel versions of the algorithm**

The parallel efficiency for the cluster system of 6 nodes is 0.81. An example (Figure 4) shows the results of running of the algorithm on cluster and on a single computer which is a part of that cluster. To build this diagram, we included in our algorithm special block which stores the maximum objective function value reached by the algorithm and the current time after each 10 steps to a special array. The horizontal line shows when the algorithm achieves the control value (65015.17). To reach this control value, the serial version of the algorithm has spent 229 minutes (13748 seconds), the parallel version has spent 59 minutes (3556 seconds). The average value (0.81) is calculated as the average speed-up coefficient after 10 runs for 5 different objective functions. It is interesting, that at the very first steps, the parallel efficiency coefficient is less than at the following steps. The possible reason of this fact is the process of parameters tuning which is performed by each node at the first steps and the intensive message passing.